\documentclass[12pt,a4paper]{article}
\usepackage{tikz,amsmath,mathtools,amssymb,amsthm}
\usepackage[margin=1.0in]{geometry}
\usepackage{cite}
\usetikzlibrary{shapes,snakes}
\usetikzlibrary{decorations.markings}
\usetikzlibrary{decorations.pathreplacing}
\usepackage[colorlinks=true
,urlcolor=blue
,anchorcolor=blue
,citecolor=blue
,filecolor=blue
,linkcolor=blue
,menucolor=blue
,linktocpage=true
,pdfproducer=medialab
,pdfa=true
]{hyperref}
\numberwithin{equation}{section}

\def\a{\alpha}
\def\b{\beta}

\def\m{\mu}

\def\be{\begin{equation}}
\def\ee{\end{equation}}
\def\bea{\begin{eqnarray}}
\def\eea{\end{eqnarray}}

\makeatletter
\renewcommand\section{\@startsection {section}{1}{\z@}%
                                   {-3.5ex \@plus -1ex \@minus -.2ex}
                                   {2.3ex \@plus.2ex}%
                                   {\normalfont\large\bfseries}}
\renewcommand\subsection{\@startsection{subsection}{2}{\z@}%
                                     {-3.25ex\@plus -1ex \@minus -.2ex}%
                                     {1.5ex \@plus .2ex}%
                                     {\normalfont\bfseries}}
\makeatother

\begin{document}
\begin{center}

	{\Large {\bf Static spherically symmetric black holes of \\ de Rham-Gabadadze-Tolley massive gravity \\in arbitrary dimensions} }
	
	\vspace*{15mm}
	\vspace*{1mm}
	{\large  Ghadir Jafari$^{a,b,}$\footnote{ghadir.jafari@mail.um.ac.ir}, M. R. Setare$^{b, c,}$\footnote{rezakord@ipm.ir} and Hamid R. Bakhtiarizadeh$^{d,}$\footnote{bakhtiarizadeh@sirjantech.ac.ir} }
	\vspace*{1cm}\\
		{ {\it $^a$School of Physics, Institute for Research in Fundamental Sciences (IPM),
			 Tehran, Iran}}\\
	{ {\it $^b$Research Institute for Astronomy and Astrophysics of Maragha (RIAAM), P.O. Box
			55134-441, Maragha, Iran}} \\
		{ {\it $^c$Department of Science, Campus of Bijar, University of Kurdistan, Bijar , Iran}} \\
	{\it $^d$Department of Physics, Sirjan University of Technology, Sirjan, Iran}
	
\vspace{20mm}

{\bf  Abstract}
\end{center}

This article is devoted to static spherically symmetric black hole solutions of dRGT (de Rham-Gabadadze-Tolley) massive gravity in the presence of cosmological constant. The unitary and non-unitary gauges are used to find the solutions in three, four and five dimensions. We show that there are two general classes of solutions. In one of them, the effect of massive potential is appeared as the effective cosmological constant. By investigating these solutions in different dimensions, we find an expression for effective cosmological constant in arbitrary dimensions.
\vfill
\newpage


\section{Introduction}

Nowadays, it is strongly believed that the universe is experiencing an accelerated expansion, and this is supported by many cosmological observations, such as SNe Ia \cite{1'}, WMAP \cite{2'}, SDSS \cite{3'} and X-ray \cite{4'}. An effective approach to calculate the current accelerating expansion of the universe is to modify the gravitational theory. One of the simplest alternative to general relativity is Brans-Dicke scalar-tensor theory \cite{1}, and amongst the most popular modified-gravity attempts, which may successfully describe the cosmic acceleration, is the $f(R)$-gravity. Here, $f(R)$ is an arbitrary function of scalar curvature \cite{2,3}. Another proposal to modify gravity suggested by Fierz and Pauli \cite{4} to equip the graviton with a non-zero mass. The massive gravitons not only may shed a light on the dark energy problem, but also could be the candidates for dark matter \cite{5,6}.

The model proposed by Fierz and Pauli is a model of non-interacting massive gravitons on a flat background. The presence of a mass term due to applying linear gravitational perturbation theory to this model, explicitly breaks the gauge invariance of general relativity. According to quantum field theory, gravitational interaction is mediated by gravitons, which are spin-$ 2 $ particles. Since the gravitational interaction is a long range force, therefore the gravitons should be massless particles. In this case, graviton has two polarizations and consequently only 2 physical degrees of freedom. On the other hand, massive graviton has 5 physical degrees of freedom. But when one extends the Fierz-Pauli theory to a curved background, the sixth degree of freedom appears which is unfortunately a ghost. This extension has been done by Boulware and Deser \cite{8}. Since the Boulware-Deser ghost field has negative kinetic energy, leads to an unstable solution in theory. de Rham and Gabadadze have shown that the ghosts fields in massive gravity can be avoided \cite{11} (see also \cite{13}). Their model is an extension of the Fierz-Pauli theory which gives the Galileon terms in the decoupling limit.\footnote{A limit in which the graviton mass $m\rightarrow 0$ and the Planck mass $M_{p}\rightarrow\infty$, but $(m^2M_{p})^{\frac{1}{3}}$ is kept fixed, known as the decoupling limit. In this limit, one can obtain an effective theory for the scalar mode in massive gravity.} From the decoupling limit data, they have presented a non-linear massive action and have shown that there is no Boulware-Deser ghost up to and including quartic order away from the decoupling limit \cite{12}.

The spherically symmetric solutions of dRGT model in four dimensions have been studied in literature. Nieuwenhuizen has obtained the Schwarzschild-de Sitter and Reissner-Nordstr\"om-de Sitter black holes as exact solutions of dRGT model \cite{14}. The authors of \cite{cs} have found static charged black hole solutions in nonlinear massive gravity. They have shown that in the parameter space of two gravitational potential parameters, below the Compton wavelength, the black hole solutions reduce to that of Reissner-Nordstr\"om via the Vainshtein mechanism in the weak field limit.
 In \cite{Koyama:2011xz} which the authors have analyzed the diagonal-metric solutions in a perturbative manner and have shown that they lead to asymptotically flat solutions. It has been observed that these solutions exhibit the vDVZ discontinuity.
 Many other interesting solutions also have been investigated, see \cite{15}-\cite{23}.

In the present paper, we are going to explore the static spherically symmetric black hole solutions of dRGT massive gravity in different dimensions. The calculations are done in the presence of cosmological constant and in the unitary and non-unitary gauges in different dimensions.  We show that, the value of cosmological constant changes due to the effect of mass terms and so we find an expression for the value of effective cosmological constant in arbitrary dimensions. 
\section{Action and the equations of motion}
We start with the following action:
\begin{equation}\label{1}
S=\frac{1}{2\kappa^2}\int d^Dx\sqrt{-g}[R+\Lambda+m^2U(g,\phi^a)],
\end{equation}
where $R$ is the Ricci scalar, $\Lambda$ is the cosmological constant and the graviton potential $ U $ expressed as the following expansion:
\begin{equation}
U(g,\phi^a)=\sum_{i=0}^{D}c_i\mathcal{U}_i(g,\phi^a),
\end{equation}
Here, $ c_i $s are constants, $\phi^a$s are four-scalar fields known as the St\"{u}ckelberg scalars introduced to restore the general covariance \cite{7} and the $ \mathcal{U}_i $s are defined by the following recursive relation:
\begin{equation}\label{Udef}
 \mathcal{U}_i=-(i-1)!\sum_{j=1}^{i}\frac{(-1)^j}{(i-j)!}\text{Tr}(\gamma^j)\mathcal{U}_{i-j}, \quad i\geq 1,
\end{equation}
with $ \mathcal{U}_0=1 $. The matrices $ \gamma $  and $ \gamma^i $ are defined as:
\begin{equation}\label{gdef}
{\gamma_{\a}}^{\m}{\gamma_{\m}}^{\b}=g^{\m\b}\partial_{\a}\phi^{a}\partial_{\m}\phi^{b}\eta_{ab},\quad {(\gamma^i)_\a}^\b={\gamma_\a}^{\m_1}{\gamma_{\m_1}}^{\m_2}\dots {\gamma_{\m_i}}^\b.
\end{equation}
 Here, $ \eta_{ab} $ is the reference metric. The gamma matrices  satisfy: $ {(\gamma^0)_\a}^\b={\delta_\a}^\b $ with  $ {(\gamma^0)_{\a\b}}=g_{\a\b} $. Furthermore, raising and lowering indices is performed by metric $ g $, so tracing can be understood by contractions, {\it i.e,} $ \text{Tr}\gamma^i=g^{\a\b}\gamma^i_{\a\b} $.\\

Using the following relations for variation of metric \cite{Hassan:2011vm,Cao:2015cti}
\begin{equation}
\delta\text{Tr}\gamma=\frac{1}{2}\gamma_{\a\b}\delta g^{\a\b},\quad \delta\text{Tr}\gamma^i=\frac{i}{2}(\gamma^i)_{\a\b}\delta g^{\a\b},
\end{equation}
one can get the equations of motion as follows:
\begin{equation}\label{Eom}
E_{\a\b}=G_{\a\b}-\frac{\Lambda}{2}g_{\a\b}+m^2X_{\a\b}=0,
\end{equation}
where $ X_{\a\b} $ is given by the expansion
\begin{equation}
X_{\a\b}=-\frac{1}{2}\sum_{i=0}^{D}c_i\left[\sum_{j=0}^{i}\frac{i!}{j!}(-1)^{i-j}\mathcal{U}_j
(\gamma^{i-j})_{\a\b}\right].
\end{equation}
Here, we set $ c_0=0 $, because we  considered the cosmological constant separately. Using the Bianchi identities also we have  the following  constraint
\begin{equation}\label{const}
\nabla^\a X_{\a\b}=0,
\end{equation}
 on the metric and scalars. Note that we use the $ \gamma $ notations instead of $ \mathcal{K} $ which is related to $ \gamma $ as: $ \mathcal{K}^\a{}_\b=\delta^\a{}_\b-\gamma^\a{}_\b $. With this notation, $U$ is defined in terms of $ \mathcal{K} $ as
\begin{equation}
U(g,\phi^a)=\sum_{i=0}^{D}\a_i\mathcal{U}_i(g,\phi^a),
\end{equation}
where $ \mathcal{U}_i $s  are defined similar to \eqref{Udef} where $\gamma$ is replaced by $\mathcal{K}$. These two notations are related to each other by a transformation between $ \a_i $ and $ c_i $ \cite{deRham:2014zqa}. Our notations keep using $ \gamma $, because we find out the solutions have a cleaner form in terms of $ c_i $, specially in $ D $ dimensions.

\section{Spherically symmetric black holes in unitary gauge}

The most general spherically symmetric ansatz for the metric in four dimensions is given by
\begin{equation} \label{sph1}
ds^2=-C(r) dt^2+A(r) dr^2+2D(r)dr dt+B(r)^2 d\Omega^2.
\end{equation}
 By using a set of coordinate transformations as: $ \tilde{t}=t+f(r) $ and $ \tilde{r}=B(r) $, one can rewrite the metric as the following diagonal form
\begin{equation}\label{sph2}
ds^2=-C(r) d\tilde{t}^2+\tilde{A}(r) d\tilde{r}^2+\tilde{r}^2 d\Omega^2,
\end{equation}
where  $ f(r) $ and $ \tilde{A}(r) $ are defined by
\begin{align}
&f=-\int \frac{D}{C}dr,\quad\tilde{A}=\frac{D^2+AC}{B^2C},
\end{align}
respectively. The unitary gauge is defined by the gauge condition $ \phi^a=\delta^a_\mu x^\mu $. But using the above coordinate transformation we will
go away from this gauge:
\begin{align} \label{gauge1}
&\phi^0\to t+f(r),\quad\phi^r\to B(r).
\end{align}
In both pictures, we are left with four independent unknown functions that must be found using field equations. There are two ways to proceed, ansatz \eqref{sph1} with scalars satisfying the unitary gauge or \eqref{sph2} which satisfy the gauge condition \eqref{gauge1}. In this section we use the former, and the latter will be applied in the next sections to find the solutions in different dimensions.

Using Eq.\eqref{gdef}, one can compute the components of matrix $ \gamma^2 $ as:
\begin{equation}
\gamma^2=\frac{1}{\Delta^2}
\left(
\begin{array}{cccc}
A & D & 0 & 0 \\
-D & C & 0 & 0 \\
0 & 0 & \frac{\Delta^2r^2}{B^2} & 0 \\
0 & 0 & 0 & \frac{\Delta^2r^2}{B^2} \\
\end{array}
\right).
\end{equation}
 Since the above matrix contains $ 2\times2 $ blocks, one can exploit the Cayley-Hamilton relation for $ 2\times2 $ matrices to find the square root of this matrix,
\begin{equation}
\text{Tr}(M) M=M^2+ {\rm I}_2(\text{Det}M),
\end{equation}
where $ {\rm I}_2 $ denotes a $ 2\times2$ identity matrix.
After some calculations, it is straightforward to find the matrix $ \gamma $, 
\begin{equation}
\gamma=\frac{1}{W \Delta }\left(
\begin{array}{cccc}
A+\Delta  & D & 0 &
0 \\
-D & C+\Delta  & 0
& 0 \\
0 & 0 & \frac{W \Delta r}{B} & 0 \\
0 & 0 & 0 & \frac{W \Delta r}{B} \\
\end{array}
\right),
\end{equation}
where $ \Delta^2=D^2+AC $ and $ W^2=A+C+2\Delta $. The eigenvalues of $ \gamma $ can be evaluated as:
\begin{equation}
\lambda_{1}=\lambda_2=\frac{r}{B},\quad \lambda_{3,4}=\frac{W\pm\sqrt{W^2-4\Delta}}{2\Delta}.
\end{equation}
Having  these results, one can easily find all components of $ \gamma^i $s and $ \mathcal{U}_i $  and thereupon different components of equations of motion
 \begin{align}\label{Ett}
 &
 E^t{}_t=\frac{1}{2 B^2}\left((2+\Lambda B^2+(\frac{B'BA}{\Delta^2})'+B''\frac{BA}{\Delta^2})\right)+\frac{m^2}{2 B^2
 	\Delta  W} \bigg\{A \left(B^2 c_1+4 B
 	c_2 r+6 c_3 r^2\right)\nonumber\\&\qquad+\Delta  \left(B^2
 	c_1+4 B
 	c_2 r+6 c_3 r^2+2 r W (B c_1+c_2r)\right)\bigg\}=0,\\
 	&E^t{}_r=\frac{ m^2D \left\{B^2 c_1+4 B c_2
 		r+6 c_3 r^2\right\}}{2 B^2 \Delta  W}+\frac{2 D
 		\left(B' \Delta '-\Delta  B''\right)}{B \Delta ^3}=0,\label{Etr}\\
 	&  E^r{}_t=-\frac{m^2 D\left(B^2 c_1+4 B c_2
 		r+6 c_3 r^2\right)}{2 B^2 \Delta  W}=0,\label{Ert}\\
 	&E^r{}_r=\frac{2 B' \left(A B'+B A'\right)}{2 B^2 \Delta
 		^2}-\frac{\left(B^2 \Lambda +2\right)}{{2 B^2}}
 	-\frac{m^2}{2 B^2
 		\Delta  W} \bigg\{C \left(B^2 c_1+4
 	B c_2 r+6 c_3 r^2\right)\nonumber\\&\qquad+\Delta
 	\left(B^2 c_1+4
 	B c_2 r+6 c_3 r^2+2 r W (B \text{c1}+c_2 r)\right)\bigg\}=0,\label{Err}\\
 	& E^\theta{}_\theta=\frac{1}{B\Delta}\left((\frac{B'A}{\Delta})'+\frac{B}{2}(\frac{A}{\Delta})'\right)+\frac{\Lambda}{2}+\frac{m^2 \{W(B c_1+2 c_2 r )+2 B c_2+c_1 \Delta  r+6 c_3 r\}}{2 B \Delta }=0.\label{Ethth}
 \end{align}
It is clear from Eq.\eqref{Ert} that there are two distinct approaches to solve this equation\footnote{This feature indicates that the known Birkhoff's theorem no more holds in these massive gravity models\cite{Koyama:2011xz}.}, which are
 \begin{align}
 D(r)=0\quad
 \text{or}\quad
 \left(B^2 c_1+4 B c_2
 r+6 c_3 r^2\right)=0.
 \end{align}
By choosing $ D=0 $, we will end with  three equations: \eqref{Ett}, \eqref{Err} and  \eqref{Ethth} which can't be solved analytically, discussion about this branch of solutions can be found in various  references\cite{16,15,Koyama:2011xz,Damour:2002gp} where some perturbative  solutions has been explored, and it is shown that the solutions exhibit the vDVZ  discontinuity. Here we choose the latter, that is
 \begin{equation}\label{Beq}
 B^2 c_1+4 B c_2
 r+6 c_3 r^2=0,
 \end{equation}
by the solution: $ B=br $. Substituting this solution into the above equation leads to an equation in terms of $ b $ which has two solutions:
 \begin{equation}
 b_{1,2}=\frac{-2
 	c_2\pm\sqrt{4 c_2^2-6 c_1 c_3}}{6 c_3}.
 \end{equation}
Moreover, from Eq.\eqref{Etr} we obtain: $ \Delta=s B' $, where $ s $ is a free constant and we choose it to be one here, in fact this constant can be removed in final solution  by rescaling of time coordinate.
 Substituting these results into Eq.\eqref{Ett} or Eq.\eqref{Err} we get the following equation:
 \begin{equation}
 rA'+A+r^2(b^2\Lambda-m^2(bc_1+c_2))-1=0,
 \end{equation}
 that has a solution as
 \begin{equation}
A=1-\frac{M}{r}+\frac{r^2}{6}(\Lambda+2bm^2(c_1+c_2b)).
 \end{equation}
On the other hand, from  Eq.\eqref{Ethth} we find that $W=1+b$.
This solution is match with the results of
\cite{Berezhiani:2011mt} for specific values of $ c_i $s. In fact, we can use the coordinate transformations stated at the beginning of this section to express the result as:
\begin{equation}
ds^2=-f(r)^2dt^2+\frac{1}{f(r)^2}dr^2+r^2d\Omega^2,
\end{equation}
where $ f^2=1-(M/r)+(r^2/6)(\Lambda+2bm^2(c_1+c_2b)) $. One can find out that the effect of massive terms is appeared in such a way that change the value of cosmological constant. However, as stated above, the coordinate transformations excite the components of St\"{u}ckelberg fields. These AdS-Schwarzschild black hole solutions with effective cosmological constant also has been found in \cite{14}, but with an approach different from our study and for specific values of constants. We will explore this type of solutions in the next section and find the value of effective cosmological constant in different dimensions.

Before closing this section, it is worth to mention that, one can use the Euler-Lagrange method to directly find the equations governing the functions $ \{W,\Delta,A,B\} $. Evaluating the lagrangian in \eqref{1} for the ansatz \eqref{sph1} we get:
\bea
{\cal L}&=&\frac{\sin (\theta)}{\Delta ^2} \left\{\Delta ^3 \left(B^2 \Lambda +2\right)+B \Delta ' \left(4 A B'+B A'\right)-\Delta  \left(B
\left(4 B' A'+B A''\right)\right.\right.\nonumber\\&&\left.\left.\qquad\quad+2 A \left(2 B B''+B'^2\right)\right)\right\}+m^2 \sin (\theta) \left\{W(B^2 c_1+4 B c_2
r+6 c_3 r^2)\right.\nonumber\\&&\left.\qquad\quad+\Delta(2c_2r^2+2c_1rB)+2c_2B^2+12c_3rB+24c_4r^2 \right\}.
\eea
As a further check, using the Euler-Lagrange equation:
\begin{equation}
\frac{ \partial {\cal L}}{\partial \phi}-\partial_r\frac{ \partial {\cal L}}{\partial \phi'}+\partial^2_r\frac{ \partial {\cal L}}{\partial \phi''}=0,
\end{equation}
with $ \phi=\{W,\Delta,A,B\} $, one can recover the previous results. For example it can be seen that, $ W $ play the role of Lagrange multiplier and its coefficient is the equation \eqref{Beq}.

 \section{Black holes in different  dimensions}
 In the previous section we used unitary gauge to find Black Hole solutions in four dimensions. In this section instead, we use non-unitary gauge to find solutions in various dimensions. The unitary calculation can be extended to other dimensions,  but we saw that this two gauges are related by coordinate transformation. So we proceed with just non-unitary gauge in order to find the result for general  dimensions, because the calculation is more tractable in this gauge.  
\subsection {3D black holes}
Here, we are going to find the spherically solutions of equation \eqref{Eom} in three dimensions and
show that the BTZ black holes are solutions of such equation. One can expect such
solution, because it is known that there is a connection
between the dGRT  formulation in three dimensions and other 3d massive gravities like NMG \cite{deRham:2014zqa,pt} , where the
BTZ is a solution. In this case the equations of motion become: 

\begin{align}
&G_{\alpha \beta} -  \tfrac{1}{2} (\Lambda+m^2U) \mathit{g}_{\alpha \beta}\nonumber\\& + \mathit{m}^2 \bigl\{\tfrac{1}{2} (\mathit{c}_1^{\text{}} + 2 \mathit{c}_2^{\text{}} \mathcal{U}_1^{\text{}} + 3 \mathit{c}_3^{\text{}} \mathcal{U}_2^{\text{}}) \gamma^{1}{}_{\alpha \beta} - ( \mathit{c}_2^{\text{}} + 3 \mathit{c}_3^{\text{}} \mathcal{U}_1^{\text{}}) \gamma^{2}{}_{\alpha \beta} + 3 \mathit{c}_3^{\text{}} \gamma^{3}{}_{\alpha \beta}\bigr\}=0.
\end{align}
We consider the following ansatz for the metric and scalars $ \phi^a $
\begin{align}
&ds^2=-f(r)^2dt^2+\frac{1}{f(r)^2}dr^2+r^2d\theta^2,\nonumber\\
&\phi^0=t+h(r),\quad\phi^i=\phi(r)\frac{x^i}{r}.
\end{align}
The components of matrix $ \gamma^2 $ can be computed using equation \eqref{gdef} as:
\begin{equation}
(\gamma^2)_\a{}^\b=
\begin{pmatrix}
&\frac{1}{f^2}&\frac{h'}{f^2}&0\\
&-f^2h'&f^2(\phi'^2-h'^2)&0\\
&0&0&\frac{\phi^2}{r^2}
\end{pmatrix},
\end{equation}
where prime denotes derivative with respect to $ r $.
The matrix $ \gamma $ then becomes:
\begin{equation}
(\gamma)_\a{}^\b=\frac{1}{\sigma}
\begin{pmatrix}
&\frac{1}{f^2}+\phi'&\frac{h'}{f^2}&0\\
&-f^2h'&f^2(\phi'^2-h'^2)+\phi'&0\\
&0&0&\frac{\sigma\phi}{r}
\end{pmatrix},
\end{equation}
where
\begin{equation}
\sigma=\sqrt{\frac{1}{f^2}+2\phi'+f^2(\phi'^2-h'^2)}.
\end{equation}
Using this result one can compute all $ \gamma^i $ matrices and consequently various components of equations of motion.
From the $ E_{01} $ component of equations of motion we get:
\begin{equation}
h'(c_1r+2c_2\phi(r))=0,
\end{equation}
which has a solution as: $ \phi(r)=b r $ with $ b=-c_1/2c_2 $.
Inserting this result into the $ E_{00} $ or $ E_{11} $ component yields to
\begin{equation}
2f'f-r(c_1bm^2+\Lambda)=0,
\end{equation}
which has the following solution
\begin{equation}
f^2=-M+\frac{r^2}{2}(c_1bm^2+\Lambda).
\end{equation}
For $  m=0 $, the above result leads to a non-rotating BTZ black hole \cite{9,10}. 
Using this  and the explicit from of $ E_{22} $ component of equations of motion, we can find the following equation for $ h $
\begin{equation}
1+f^4(b^2-h'^2)=(1+b^2)f^2,
\end{equation}
solving this equation in terms of  $h$ yields
\begin{equation}\label{heq}
h(r)=\pm\int_{1}^{r}\frac{\sqrt{(f(z)^2-1)(b^2f(z)^2-1)}}{f(z)^2}dz.
\end{equation}
These solutions also satisfy the constraint \eqref{const}. We also can obtain equation \eqref{heq} by using the $ r $ component of this constraint. We see that the constant $ c_3 $ does not appear in the solution functions, this reflect the fact that $ \mathcal{U}_3 $ is a total derivative
\begin{equation}
\mathcal{U}_3=6\phi\phi'=3(\phi^2)'.
\end{equation}
It is also notable that, by inserting the following ansatz
\begin{align}
&ds^2=-f(r)^2dt^2+g(r)^2dr^2+r^2d\theta^2,\nonumber\\
&\phi^0=t+h(r),\quad\phi^i=\phi(r)\frac{x^i}{r},
\end{align}
into the lagrangian and finding the equations of motion for $ f(r) $, $ g(r) $, $ h(r) $ and $ \phi(r) $, we exactly restore the previous results.
\subsection {4D black holes}
In this subsection, we search for the static spherically symmetric black hole solutions in 4 dimensions. In this case, the equations of motion become:
\begin{align}
&G_{\alpha \beta} -  \tfrac{1}{2} (\Lambda+m^2U) \mathit{g}_{\alpha \beta} + \mathit{m}^2 \bigl\{\tfrac{1}{2} (\mathit{c}_1^{\text{}} + 2 \mathit{c}_2^{\text{}} \mathcal{U}_1^{\text{}} + 3 \mathit{c}_3^{\text{}} \mathcal{U}_2^{\text{}} + 4 \mathit{c}_4^{\text{}} \mathcal{U}_3^{\text{}}) \gamma^{1}{}_{\alpha \beta}\nonumber\\& - ( \mathit{c}_2^{\text{}} + 3 \mathit{c}_3^{\text{}} \mathcal{U}_1^{\text{}} + 6 \mathit{c}_4^{\text{}} \mathcal{U}_2^{\text{}}) \gamma^{2}{}_{\alpha \beta} + 3 (\mathit{c}_3^{\text{}} + 4 \mathit{c}_4^{\text{}} \mathcal{U}_1^{\text{}}) \gamma^{3}{}_{\alpha \beta} - 12 \mathit{c}_4^{\text{}} \gamma^{4}{}_{\alpha \beta}\bigr\}=0.
\end{align}
Similar to the 3-dimensional case, we suppose the following ansatz for the metric and the St\"{u}ckelberg scalars
\begin{align}
&ds^2=-f(r)^2dt^2+\frac{1}{f(r)^2}dr^2+r^2d\Omega^2,\nonumber\\
&\phi^0=t+h(r),\quad\phi^i=\phi(r)\frac{x^i}{r}.
\end{align}
Using the above ansatz, the equation of $ \phi $ becomes
\begin{equation}
(c_1r^2+4c_2r\phi+6c_3\phi^2)=0.
\end{equation}
The solution is: $ \phi=b r $, where $ b $ is the roots of the second-order equation:
$ c_1+4c_2b+6c_3b^2 =0$, that are
\begin{equation}
b_{1,2}=\frac{-2
	c_2\pm\sqrt{4 c_2^2-6 c_1 c_3}}{6 c_3}.
\end{equation}
Also, the equation for $ f $ takes the following form
\begin{equation}
2rf'f+f^2-\frac{r^2}{2}(\Lambda+2bm^2(c_1+c_2b))-1=0.
\end{equation}
The solution of the above equation is given by
\begin{equation}
f^2=1-\frac{M}{r}+\frac{r^2}{6}(\Lambda+2bm^2(c_1+c_2b)).
\end{equation}
For $ m=0 $, the metric reduces to the usual AdS-Schwarzschild black hole.
 The equation for $ h $ is similar to 3D case:
\begin{equation}\label{heqtion}
1+f^4(b^2-h'^2)=(1+b^2)f^2.
\end{equation}
So the solution is the same as Eq.\eqref{heq}. With this result, the constraint \eqref{const} is satisfied, as well.
\subsection {5D black holes}
Here, the equations of motion become:
\begin{align}
&G_{\alpha \beta} -  \tfrac{1}{2} (\Lambda+m^2U) \mathit{g}_{\alpha \beta} + \mathit{m}^2 \bigl\{\tfrac{1}{2} (\mathit{c}_1^{\text{}} + 2 \mathit{c}_2^{\text{}} \mathcal{U}_1^{\text{}} + 3 \mathit{c}_3^{\text{}} \mathcal{U}_2^{\text{}} + 4 \mathit{c}_4^{\text{}} \mathcal{U}_3^{\text{}} + 5 \mathit{c}_5^{\text{}} \mathcal{U}_4^{\text{}}) \gamma^{1}{}_{\alpha \beta} \nonumber\\&- ( \mathit{c}_2^{\text{}}+ 3 \mathit{c}_3^{\text{}} \mathcal{U}_1^{\text{}} + 6 \mathit{c}_4^{\text{}} \mathcal{U}_2^{\text{}} + 10 \mathit{c}_5^{\text{}} \mathcal{U}_3^{\text{}}) \gamma^{2}{}_{\alpha \beta} + 3 (\mathit{c}_3^{\text{}} + 4 \mathit{c}_4^{\text{}} \mathcal{U}_1^{\text{}} + 10 \mathit{c}_5^{\text{}} \mathcal{U}_2^{\text{}}) \gamma^{3}{}_{\alpha \beta}\nonumber\\& - 12 (\mathit{c}_4^{\text{}} + 5 \mathit{c}_5^{\text{}} \mathcal{U}_1^{\text{}}) \gamma^{4}{}_{\alpha \beta} + 60 \mathit{c}_5^{\text{}} \gamma^{5}{}_{\alpha \beta}\bigr\}=0.
\end{align}
Consider the ansatz
\begin{equation}
ds^2=-f(r)^2dt^2+\frac{1}{f(r)^2}dr^2+r^2d\Omega_3^2,
\end{equation}
where $ d\Omega_3 $ is the metric on the unit three sphere. In 5-dimensional case, we have a third-order equation for $ \phi $:
\begin{equation}
(c_1r^3+6c_2r^2\phi+18c_3r\phi^2+24c_4\phi^3)=0.
\end{equation}
Taking the solution as: $ \phi=br $, one obtains the following equation for the constants $c_i$:
\begin{equation}
c_1+6b(c_2+b(3c_3+4bc_4))=0.
\end{equation}
The equation governing $ f $ becomes:
\begin{equation}
r f(r)
f'(r)+f(r)^2-\frac{r^2}{6} \left(\Lambda+ 3 \mathit{b} \mathit{m}^2 (c_1+2 \mathit{b} (c_2+\mathit{b} c_3)) \right)-1=0,
\end{equation}
which has a solution as:
\begin{equation}
f^2=1-\frac{2M}{r^2}+\frac{r^2}{12}\left\{\Lambda+ 3 \mathit{b} \mathit{m}^2 (c_1+2 \mathit{b} (c_2+\mathit{b} c_3)) \right\}.
\end{equation}
Also, it is interesting that, there is an  equation for $ h $ similar to the 3 and 4-dimensional cases.
\subsection {Black holes in D dimensions}
As can be seen from the previous sections, it seems that the value of effective cosmological constant in different dimensions can be written as a series in $ b $, where $ b $ is a coefficient in the solution of $ \phi $, which always has the form: $ \phi=br $. We also have explored the form of effective cosmological constant in six dimensions and find out that the previous pattern is repeated:

\begin{equation}
	\Lambda_{eff}=\Lambda+4bm^2\left\{c_1+3b(c_2+2b(c_3+bc_4))\right\}.
\end{equation}
Putting the results in different dimensions together, one may guess the following pattern for the effective cosmological constant in general dimensions $ D $:
\begin{equation}
	\Lambda_{eff}=\Lambda+(D-2)bm^2\left\{c_1+(D-3)b(c_2+(D-4)b(c_3+(D-5)b(c_4+...)))\right\},
\end{equation}
which can be rewritten in the following compact form:
\begin{equation}
	\Lambda_{eff}=\Lambda+m^2\sum_{i=1}^{D-2}\frac{(D-2)!}{(D-2-i)!}c_ib^i,
\end{equation}
where $ b $ is one of the roots of the following equation:
\begin{equation}
	\sum_{i=1}^{D-1}\frac{i(D-2)!}{(D-i-1)!}c_i b^{i-1}=0.
\end{equation}
As a further check, using the standard Mathematica package ``xAct" \cite{xact,Nutma:2013zea}, we validate the above results even for seven and eight dimensions and find an exact agreement.  

\subsection{Charged solutions}
In this section, we search for charged black hole solutions. For this purpose, we add the Maxwell term to the action (\ref{1}) 
\begin{equation}
S=\frac{1}{2\kappa}\int d^Dx\sqrt{-g}[R+\Lambda+\frac{1}{4}F_{\mu\nu}F^{\mu\nu}+m^2U(g,\phi^a)].
\end{equation}
We consider the gauge field to be as
\begin{equation}
A_\mu=a(r)dt.
\end{equation}
Using the Maxwell equations, we obtain the following experssions for $a(r)$ in 3 and 4-dimensional cases
\begin{equation}
a(r) =
\left\{
\begin{array}{ll}
Q\text{Log}(r)  & \mbox{in 3D} \\
\frac{Q}{r} & \mbox{in 4D}
\end{array}
\right.
\end{equation}
Also, the function $ f(r) $ takes the following form in 3, 4 and 5-dimensional cases, respectively:
\begin{align}
&f^2=-M+\frac{r^2}{2}(\Lambda+c_1bm^2)+\frac{Q^2}{2}\text{Log}(r),\quad\nonumber\\&
f^2=1-\frac{M}{r}+\frac{r^2}{6}(\Lambda+2bm^2(c_1+c_2b))-\frac{Q^2}{4r^2},\quad\nonumber\\&
f^2=1-\frac{2M}{r^2}+\frac{r^2}{12}\left\{\Lambda+ 3 \mathit{b} \mathit{m}^2 (c_1+2 \mathit{b} (c_2+\mathit{b} c_3)) \right\}-\frac{Q^2}{12r^4}.\quad
\end{align}
So again we see that the effect of massive potential appears as effective cosmological constant, independent of the charge. 

\section{Summary and discussion}\label{dis}

We have studied the static spherically symmetric black hole solutions of dRGT massive gravity in the presence of cosmological constant in various dimensions. It is shown that the presence of mass term  change the value of cosmological constant and unlike some claims, does not add power-like $ r^{-\lambda} $ or logarithmic terms $ \text{Log}(r) $ to $ f(r) $ \cite{Li:2016fbf}.

We have found that the value of effective cosmological constant in different dimensions is a series in terms of $ b $, where $ b $ is appeared as a coefficient in the solution of $ \phi $, which always has the form: $ \phi=br $. The explicit form of effective cosmological constant in general dimensions $ D $ is:
\begin{equation}
\Lambda_{eff}=\Lambda+m^2\sum_{i=1}^{D-2}\frac{(D-2)!}{(D-2-i)!}c_ib^i,
\end{equation}
where $ b $ is one of the roots of the following equation:
\begin{equation}
\sum_{i=1}^{D-1}\frac{i(D-2)!}{(D-1-i)!}c_i b^{i-1}=0.
\end{equation}
From the above results, it is deduced that in $ D $ dimensions there are $ D-2 $ different vacuum solutions. From this point of view, the result is similar to vacuum solutions of Lovelock Gravity, for example see \cite{Camanho:2015ysa}.
  So in general the solutions can be asymptotic AdS, dS or flat solutions. The theory exhibits degenerate behavior whenever two or more effective
  cosmological constants coincide.
  
For a charged black hole, the effect of charge is appeared as an extra term in $ f(r) $ and decoupled from mass terms. Therefore, the above expression for effective cosmological constant is satisfied anyway, even for a charged black hole. 

\section*{Acknowledgement}\addcontentsline{toc}{section}{Acknowledgement}
We would like to thank Gregory Gabadadze for careful reading of the manuscript and helpful comments.
We also thank the anonymous referee for important comments. 
This work has been supported financially by Research Institute for Astronomy \& Astrophysics of Maragha (RIAAM) under the project No.1/4717-161.

\appendix
\section{The Basic Invariant $ I^{ab} $}

In addition to the usual ones in GR, due to the existence of St\"{u}ckelberg fields, there is another basic invariant in massive gravity which is defined by: $ I^{ab}=g^{\mu\nu}\partial_{\mu}\phi^{a}\partial_{\nu}\phi^{b} $. In  
\cite{Berezhiani:2011mt} de Rham and his colleagues have pointed out that the singularities  in $ I^{ab} $ are problematic for the existence  of fluctuations around the classical
solutions exhibiting it. Therefore, in order for the solutions to be true, the coordinate singularities, like those one usually appear at the horizon, must be absent in $ I^{ab} $. In this appendix we are going to show that these singularities are really absent in the solutions found in this paper in arbitrary dimensions. 

For the general ansatz :
\begin{align}
&ds^2=-f(r)^2dt^2+\frac{1}{f(r)^2}dr^2+r^2d\Omega^2_{D-2},\nonumber\\
&\phi^0=t+h(r),\quad\phi^i=\phi(r)\frac{x^i}{r},
\end{align}
the $ I^{ab} $ can be calculated as:
\begin{align}
&I^{00}=f^2 h'^2-\frac{1}{f^2},\nonumber\\
&I^{0i}=I^{i0}=f^2\phi'h'n^{i},\nonumber\\
&I^{ij}=\frac{\phi^2}{r^2}\delta^{ij}+(f^2\phi'^2-\frac{\phi^2}{r^2})n^in^j,
\end{align}
where $ n^i=\frac{x^i}{r} $. Using $ \phi=br $ and the equation \eqref{heqtion}, which is general
 for all dimensions, we obtain:
\begin{equation}\label{key}
I^{00}=f^2b^2-(1+b^2).
\end{equation}
The expression for $ f $ in general dimensions is:
\begin{equation}\label{key}
f^2=1-\frac{2M}{r^{D-3}}+\frac{\Lambda_{eff}}{(D-2)(D-1)}r^2+\frac{Q^2}{r^{2(D-3)}},
\end{equation}
so we see that the expression for $ I^{00} $ won't become singular except at physical singularity $ r=0 $. 
Similarly for $ I^{0i} $ and $ I^{ij} $ we have:
\begin{align}
&I^{0i}=b\sqrt{(f^2-1)(b^2f^2-1)} n^i,\nonumber\\
&I^{ij}=b^2\delta^{ij}+b^2(f^2-1)n^in^j,
\end{align}
which also just become singular at physical singularity $ r=0 $, because coordinate singularities appear only in the
 negative powers of $ f $.

\providecommand{\href}[2]{#2}\begingroup\raggedright

\endgroup
\end{document}